\documentclass[a4paper,sn-mathphys-num,twocolumn]{sn-jnl}

\usepackage{geometry}

\usepackage[version=4]{mhchem}

\usepackage{graphicx}
\usepackage{amsmath,amssymb,amsfonts}
\usepackage{amsthm}
\usepackage{mathrsfs}
\usepackage[title]{appendix}
\usepackage{xcolor}
\usepackage{textcomp}
\usepackage{manyfoot}
\usepackage{algorithm}
\usepackage{algorithmicx}
\usepackage{algpseudocode}
\usepackage{listings}
\usepackage{caption}
\usepackage{booktabs}
\usepackage{multirow}
\usepackage{tabularx}
\usepackage{array}

\theoremstyle{thmstyleone}%
\theoremstyle{thmstyletwo}%

\theoremstyle{thmstylethree}%

\begin{document}

\title[Co layer Thickness Driven Structural Evolution and Magnetic Behaviour of Co/Ir/Co Trilayers]{Co layer Thickness Driven Structural Evolution and Magnetic Behaviour of Co/Ir/Co Trilayers}

\author[1]{\fnm{Subhavenkadesh} \sur{SS}}
\author[1]{\fnm{Dushyant} \sur{Singh}}
\author[1]{\fnm{Tharundev} \sur{VV}}
\author[1]{\fnm{Subha} \sur{Maity}}
\author[1]{\fnm{Navneet} \sur{Sharma}}
\author[1]{\fnm{Vigneshwari} \sur{SR}}
\author*[1]{\fnm{Krista} \sur{R. Khiangte}}\email{krista.khiangte@iitgn.ac.in}

\affil[1]{\orgdiv{Department of Physics}, \orgname{Indian Institute of Technology Gandhinagar}, \orgaddress{\street{Palaj}, \state{Gujarat}, \postcode{382055},\country{India}}}


\abstract{Ferromagnetic heavy-metal (FM/HM) trilayer heterostructures provide a versatile platform for investigating the interplay between structural and magnetic properties. In this study, we investigate the effect of Co-layer thickness on the structural and magnetic properties of sputter-deposited $\text{Co}(t_{\text{Co}})/\text{Ir}(5\text{nm})/\text{Co}(t_{\text{Co}})$ varying from 2 to 10 nm. Structural characterization using grazing-incidence X-ray diffraction (GIXRD) and high-resolution transmission electron microscopy (HRTEM) reveals a nanocrystalline structure, with the crystallite size increasing with Co-layer thickness. Furthermore, X-ray reflectivity (XRR) reveals a thickness-dependent variation in interface morphology, with the interface roughness ratio reaching a maximum at $t_{Co}$= 4 nm. Temperature-dependent zero-field-cooled and field-cooled (ZFC–FC) magnetization measurements reveal thickness-dependent characteristic temperatures and multiple relaxation regimes in selected samples, indicating complex thermally activated magnetic behaviour. In contrast, field-dependent hysteresis measurements reveal that the saturation magnetization increases with Co-layer thickness, whereas the coercivity exhibits a pronounced non-monotonic variation, with enhanced values at intermediate thicknesses. Moreover, low-temperature hysteresis loops display thickness-dependent exchange-bias-like shifts, particularly for $t_{Co}$ = 2–4 nm. These findings highlight the strong influence of Co-layer thickness on the structural and magnetic properties of Co/Ir/Co trilayers.
}

\keywords{Trilayer, Interfacial magnetism, Blocking temperature, Exchange Bias Effect, Thermally activated reversal}



\maketitle

\section{Introduction}\label{sec1}
Magnetic heterostructures consisting of alternating ferromagnetic (FM: Ni, Co, Fe, NiFe, etc.,) and non-magnetic (NM) layers exhibit a range of useful magnetic and magnetotransport properties, including giant magnetoresistance  \cite{Santamaria2026, Baibich1988, Chappert2007}, exchange bias (EB)\cite{Wang2002, Pham2026}, interlayer exchange coupling \cite{Tryggvason2026, Zhou2023}, anomalous Hall effect \cite{Sudhansu2014, singh2025viable}, anisotropic magnetoresistance  \cite{Avci2015, singh2020}, and perpendicular magnetic anisotropy \cite{Lonsky2022, Johnson1996, Singh2022deadlayerstatic}. However, FM/NM structures employing metallic spacer layers can suffer from current-shunting effects due to the high conductivity of the NM metal, which can suppress the measured magnetotransport response \cite{Xu2008, Roy2023, DSINGH2024171975}. To address this limitation, semiconductor spacers such as Si \cite{Grundy2000, Liu2007, Singh2022deadlayerstatic}, Ge \cite{Lucinski2011}, and GaAs \cite{Bensch2001} have been incorporated into FM/NM heterostructures. FM/semiconductor structures offer additional advantages, including tunable $T_C$ near ambient conditions \cite{singh2025viable}, compatibility with semiconductor processing, reduced current shunting \cite{Xu2008, Roy2023, DSINGH2024171975}, and high Hall sensitivity \cite{Das2015}. However, their magnetic and magnetotransport responses remain constrained by the weak intrinsic spin–orbit interaction (SOI) of semiconductor materials. In this context, replacing the NM semiconductor spacer with a heavy-metal (HM: Ta, W, Ir, Pt, etc.,) layer provides an effective route to further enhance magnetic and magneto-transport phenomena \cite{Tang2018, Han2021, Xinyu2020}. Compared with conventional FM/NM-metal and FM/NM-semiconductor heterostructures, these strong SOI and interfacial effects enable enhanced anomalous Hall effect (AHE) and sign reversal of AHE \cite{Zhang2023},  perpendicular magnetic anisotropy (PMA) \cite{Lau2019}, and larger spin–orbit torque (SOT) generated via the large spin Hall effect (SHE) facilitating efficient magnetisation switching,  interfacial Dzyaloshinskii-Moriya interaction (DMI) \cite{Fan2023}, and current-driven auto-oscillations in the FM layer \cite{Akash2023}. Leveraging these enhanced Spin Orbit Coupling (SOC) mechanisms is essential for next-generation technological applications, including high-density magnetic recording, energy-efficient quantum computing, high-frequency magnetisation dynamics, magnetic memory, sensors, logic devices and advanced spin-orbitronic devices \cite{Dieny2020}.

Among Ta, W, Ir, and Pt, Ir is particularly attractive when combined with FM layers because it provides strong spin–orbit interaction (SOI) \cite{Dutta2021, Ma2020} and can induce strong interfacial DMI \cite{Saito2022}. Ir inserted in between two FM layers can also induce antiferromagnetic interlayer coupling \cite{Fan2023}, representing another way to tune the magnetics of multilayer structures for spintronic applications. In particular, the Ir/Co combination is well suited for enhancing interlayer coupling, perpendicular magnetic anisotropy (PMA), and interfacial DMI \cite{Ishikuro2019, Dutta2021, Ma2020}. While Co/Ir heterostructures have been extensively studied for SOT, DMI, PMA, AHE sign reversal and interlayer coupling, a systematic understanding of the correlation between $t_{\text{Co}}$ and the structural and interfacial evolution with temperature-dependent magnetic behaviour in Co/Ir/Co trilayers is still lacking. The interlayer exchange coupling via Ir oscillates and decays very fast with the spacer thickness and is only relevant for Ir layers below $\sim 1$ nm\cite{Parkin1991, Fan2023}. The difference between layers: bottom Co layer grows on HF-etched Si, where Co–Si intermixing is expected \cite{Roy2023}, top layer grows on polycrystalline Ir and is exposed to air. The result is a structure that combines Co/Ir interfaces of interest for spin-orbit devices with two inequivalent Co layers, whose properties can be compared as their thicknesses are varied. This is especially important heavy-metal layer with strong SOI but also an interlayer that can mediate magnetic coupling between the two Co layers, making Co/Ir/Co a model system for studying a model system to study the interaction between interface-driven and thickness-dependent magnetic phenomena. Thus, a considerable lack of systematic correlation between $t_{\text{Co}}$-dependent microstructural evolution and saturation magnetisation ($M_s$), coercivity ($H_c$), blocking temperature ($T_b$) and magnetic phases is observed. In particular, by measuring the zero-field cooling (ZFC) and field-cooling (FC) magnetisation as a function of temperature, we can observe the blocking and unblocking of magnetic spins and understand the thermally activated spin dynamics and magnetic phase evolution. The formation of such correlations is essential to understand and control the magnetic properties and tunability of Co/Ir trilayers for spintronic applications.

\section{Experimental details}\label{sec2}

A series of trilayer heterostructures of the form Co$(t_{Co})$/Ir$(t_{Ir})$/Co$(t_{Co})$ were deposited on Si(111)-oriented substrates using a DC/RF magnetron sputtering system. Here, $t_{Co}$ and $t_{Ir}$ denote the nominal thicknesses of the Co and Ir layers, respectively. Commercially available Co and Ir targets with a purity of 99.95\% were used to deposit the Co and Ir layers, respectively. In this series, the thickness of the iridium layer was fixed at $t_{Ir}$ = 5 nm, and the thickness of the Co layer was varied to 2, 3, 4, 6, and 10 nm. Prior to the actual deposition, the sputtering chamber was evacuated to a base pressure of approximately $2.0\times10^{-6}$~mbar to minimise contamination from residual gases and other impurities. The deposition was carried out in an Ar plasma atmosphere at a pressure of 4.0$\times$10$^{-3}$ mbar. Before loading the Si(111) substrates into the vacuum chamber, they were subjected to a systematic cleaning procedure using acetone, isopropyl alcohol (IPA), deionized (DI) water, and high-purity nitrogen (N$_2$). First, ultrasonically cleaned for approximately 10~min using a sequential IPA--acetone--IPA to remove organic contaminants and surface residues from the substrate surface. The cleaned Si(111) substrates were subsequently etched in a {2\%} HF solution for 15~s to remove the native SiO$_2$ layer from the surface. Following HF etching, the substrates were thoroughly rinsed with DI water and dried using high-purity N$_2$. The cleaned Si (111) substrate was then promptly loaded into the vacuum chamber, which was evacuated to the required base pressure, followed by three cycles of purging with high-purity Ar gas to minimise residual contaminants inside the chamber.

The structural properties of the prepared samples were characterised using a high-resolution X-ray diffraction (HRXRD) instrument of Rigaku SmartLab operated at an accelerating voltage of 30~kV, with Co $K_{\alpha}$ radiation ($\lambda = 1.7906~\text{\AA}$)  as the X-ray source. The microstructure of the samples was investigated using high-resolution transmission electron microscopy (HRTEM) of FEI, Themis 60-300 equipped with an EDS detector, and FEI-Ceta 4k $\times$ 4k camera. X-ray photoelectron spectroscopy (XPS) measurements were performed using a Thermo Scientific K-Alpha system equipped with a monochromated Al K$\alpha$ X-ray source, a 180\text{\textdegree} hemispherical analyser, dual-beam charge compensation, and depth-profiling capabilities. Depth profiling was performed using Ar$^+$ ion sputtering to investigate the depth-dependent chemical states and distributions of Co, Ir, and O within the trilayer structure. The magnetic field-dependent magnetisation measurements were carried out using a magnetometer (Cryogenic Limited, UK) Physical Properties Measurement System (PPMS) with the field applied parallel to the sample plane. The MH loops were measured at the selected temperatures after zero-field cooling from room temperature to the measurement temperature within the applied-field range.

\section{Results and Discussion}\label{sec3}

\subsection{Structural and microstructural Study}\label{subsec2}

\begin{figure}[t]
    \centering
    \includegraphics[width=1.0\linewidth]{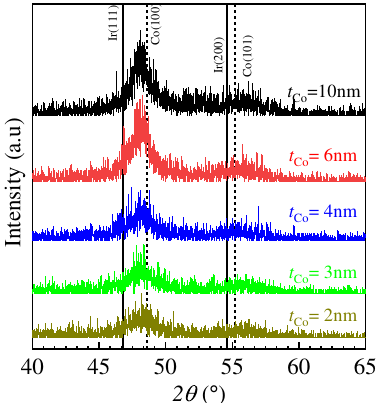}
    \caption{ GIXRD patterns of the $\text{Co}(t_{\text{Co}})/\text{Ir}(5\,\text{nm})/\text{Co}(t_{\text{Co}})$ trilayers. The diffraction peaks corresponding to bulk Co and Ir are indicated by vertical dashed and solid lines, respectively, and are indexed according to the standard JCPDS cards No. 05-0727 for Co \cite{Zheng2008} and 06-0598 for Ir \cite{Puthalath2026}.}
    \label{fig:GIXRD}
\end{figure}

Establishing the crystallographic structure is essential for interpreting the subsequent microstructural and magnetic properties, as crystal structure and texture can strongly influence the magnetic behaviour of the trilayers. Therefore, grazing-incidence X-ray diffraction (GIXRD) was first employed to investigate the crystallographic structure of the Co/Ir/Co trilayers using an HRXRD instrument, as conventional out-of-plane XRD measurements were dominated by the substrate contribution and did not reveal discernible diffraction peaks from the trilayer structure. The room-temperature GIXRD profiles of $\text{Co}(t_{\text{Co}})/\text{Ir}(5\,\text{nm})/\text{Co}(t_{\text{Co}})$ trilayers, recorded over the $2\theta$ range of $20^\circ$--$90^\circ$, are depicted in Fig. \ref{fig:GIXRD}.It is evident from the Fig. \ref{fig:GIXRD} that all the samples exhibit a broad peak centred around $48^\circ$-$50^\circ$ and a weak diffraction peak near $55^\circ$. The broad diffraction features indicate the nanocrystalline nature of the sputter-deposited Co/Ir/Co trilayers.

The diffraction features observed in the $2\theta$ range of $48^\circ$--$50^\circ$ arise from the overlap of multiple crystallographic reflections. Indexing the diffraction features using the standard JCPDS cards 06-0598 for Ir \cite{Puthalath2026} and No. 05-0727 for Co \cite{Zheng2008} indicates that the prominent broad feature in this region originates from the superposition of the face-centred cubic (FCC) Ir(111) and hexagonal close-packed (HCP) Co(100) reflections. For crystallites of about 5~nm the expected peak width is close to 2$^{\circ}$, larger than the separation between these reflections, so they appear as one broad feature. This overlap arises from the comparable interplanar spacings of these crystallographic planes, resulting in closely spaced diffraction positions when measured using Co-K$\alpha$ radiation. Consequently, the individual reflections are not well resolved and appear as a single broadened diffraction feature. Similarly, the weak diffraction feature observed near $55^\circ$ can be attributed to the overlapping FCC-Ir(200) and HCP-Co(101) reflections. Similar to the present GIXRD results, the superposition of the Co(100), Co(002), and Co(101) diffraction peaks has also been reported in Si/Co multilayers\cite{Roy2023}. To quantify the structural evolution with Co-layer thickness, an effective crystallite size ($D$) was estimated from the width of the main diffraction feature using the Scherrer equation \ref{eq:C_Size} \cite{Scherrer1918, DSINGH2024171975, singh2020}.

\begin{figure}[t]
    \centering
    \includegraphics[width=1\linewidth]{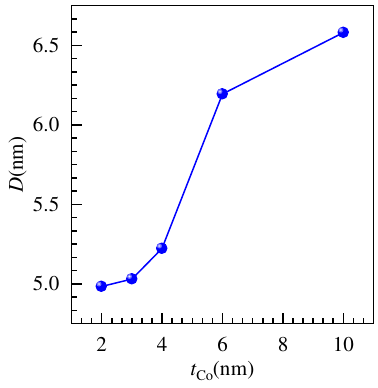}
    \caption{Crystallite size D of the {Co$(t_{Co})$/Ir(5nm)/Co$(t_{Co})$} trilayers as a function of $t_{Co}$.}
    \label{fig:C_Size}
\end{figure}

\begin{equation}
    D = \frac{K \lambda}{\beta \cos\theta}
    \label{eq:C_Size}
\end{equation}
Where $D$ is the average crystallite size, $\lambda$ denotes the wavelength of Co-K$\alpha$ X-rays used ($\lambda = 1.7906$~\AA), $\beta$ is the full width at half maximum of the peak, $\theta$ is the Bragg angle, and $K$ is the so-called Scherrer constant. In general, $K$ is affected by the shape and size distribution of crystallites, the indices of diffraction lines, etc. The value of $K$ ranges from 0.62 to 2.08 \cite{Singh2022deadlayerstatic}. In the present calculation, $K$ = 0.91 is used \cite{Langford1978, singh2020}. The estimated $D$ value of the nanocrystallites is presented in Fig. \ref{fig:C_Size}. Apparently, the crystallite size decreases from approximately 6.5 nm for the $t_{Co}$= 10 nm sample to approximately 4.98 nm for the $t_{Co}$= 2 nm sample. For the thinnest samples, the Ir spacer makes up most of the scattering volume, so the value there largely reflects the Ir grains; the increase with $t_{\mathrm{Co}}$ shows that thicker Co layers contain laterally larger coherent domains. This evolution indicates that changing the cobalt thickness changes the microstructural length of the magnetic trilayer. The variation in crystallite size is particularly relevant to the magnetic response because, in a nanocrystalline ferromagnetic layer, the crystallite size, grain boundaries, and local variations in crystallographic orientation can significantly influence the magnetic anisotropy, saturation magnetisation, blocking temperature, and magnetisation reversal processes \cite{singh2025viable, Herzer1990}.

\begin{figure}[t!]
    \centering
    \includegraphics[width=1\linewidth]{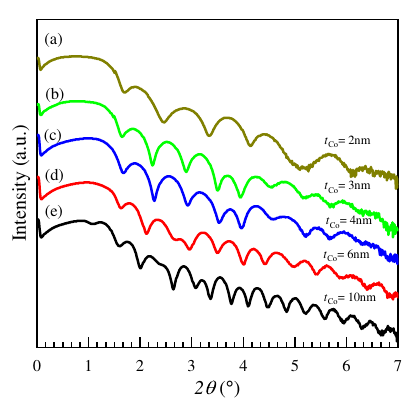}
    \caption{XRR patterns of the {Co$(t_{Co})$/Ir(5nm)/Co$(t_{Co})$} trilayers recorded at 300 K for various Co layer thicknesses.}
    \label{fig:XRR}
\end{figure}

\begin{figure}[b!]
    \centering
    \includegraphics[width=1\linewidth]{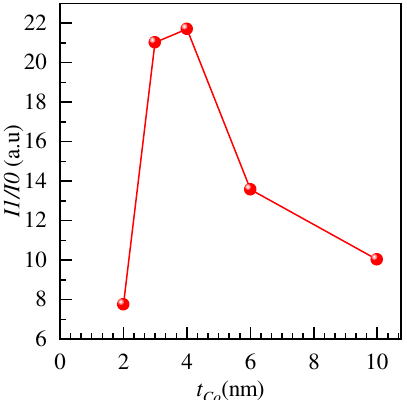}
    \caption{Ratio $I_{1}/I_{0}$ of the second- to the first-order XRR fringe maximum vs $(t_{Co})$.}
    \label{fig:I}
\end{figure}

Additionally, X-ray reflectivity (XRR) measurements were performed to evaluate the interface quality of the Co/Ir/Co trilayers as a function of Co-layer thickness. Unlike diffraction techniques, XRR probes the electron-density profile normal to the film plane, giving information on the layer thickness and the interfacial morphology, which are relevant for the understanding of the magnetic properties mediated by the Co/Ir interfaces. The XRR patterns of the \(\text{Co}(t_{\text{Co}})/\text{Ir}(5\,\text{nm})/\text{Co}(t_{\text{Co}})\) trilayers, with the corresponding \(t_{\text{Co}}\) values indicated against each pattern, are shown in Fig. \ref{fig:XRR}. Well-resolved Kiessig fringes are observed in the XRR patterns of all the samples. These subsidiary oscillations arise from the interference of X-rays reflected from the air–film and film–substrate interfaces \cite{singh2025viable, Parratt1954}. As the Co-layer thickness increases, the spacing between consecutive Kiessig fringes gradually decreases, while the number of observable oscillations increases. This behaviour confirms the expected increase in the overall trilayer thickness as Co-layer thickness increases and is consistent with the inverse relationship between fringe periodicity and film thickness. Further, to investigate the evolution of interface quality with $t_{Co}$, the intensity ratio of the second-order peak ($I_{1}$) to the first-order peak ($I_{0}$) in the XRR kiessing fringes was evaluated and reported elsewhere \cite{singh2025viable}. The percentage intensity ratio $I_{1}/I_{0}$ is plotted as a function of co-layer thickness in Fig. \ref{fig:I}. The intensity ratio initially increases with increasing $t_{Co}$, reaches a maximum at $t_{Co}=4$ nm, and subsequently decreases as $t_{Co}$ increases further. Based on the adopted intensity-ratio criterion, a higher $I_{1}/I_{0}$ ratio indicates rougher interface quality, with higher interface roughness for samples $t_{Co}$ = 4 nm than in other samples. Therefore, the interface roughness is inferred to be highest for the sample with $t_{Co}$ = 4 nm.

\begin{figure}[b]
    \centering
    \includegraphics[width=1.0\linewidth]{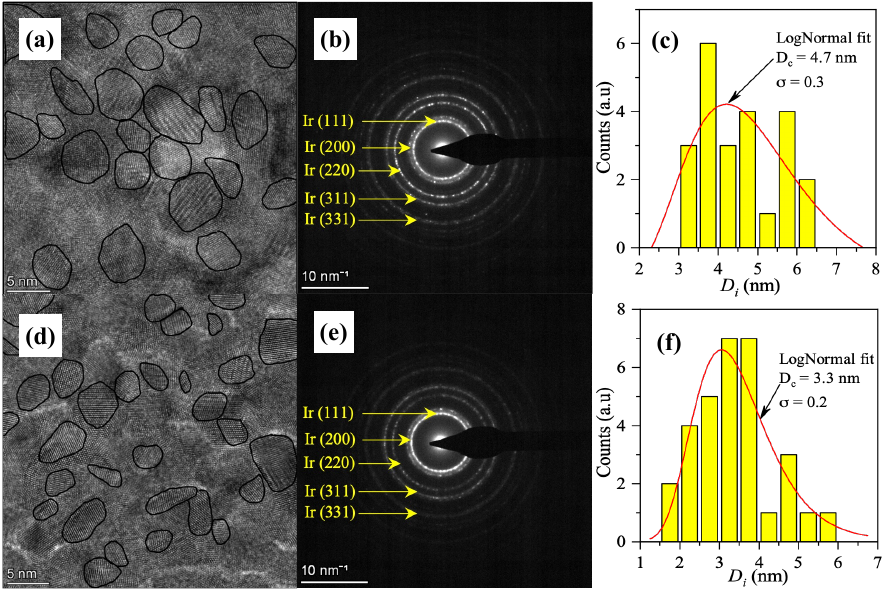}
    \caption{HRTEM images of the {Co$(t_{Co})$/Ir(5nm)/Co$(t_{Co})$} trilayers with (a) $t_{Co}$= 6 nm and (d) $t_{Co}$= 2 nm. The solid curves are drawn to show the crystalline boundaries. (b) and (e) are the corresponding SAED patterns of the HRTEM  images depicted in (a) and (d), respectively. (c) and (f) distribution of nanocrystallite sizes fitted with a lognormal distribution equation for $t_{Co}$ = 6 nm and 2 nm.}
    \label{fig:TEM}
\end{figure}

Since the Co/Ir/Co trilayers have thicknesses of only a few nanometres, further microstructural characterisation using HRTEM combined with selected area electron diffraction (SAED) is essential to corroborate the polycrystalline structural phases identified from the GIXRD analysis and to examine the distribution of the nanocrystallites, which are important for understanding the temperature-dependent and field-dependent magnetisation behaviour and the associated magnetic parameters. For this purpose, two selected $\mathrm{Co}(t_{Co})/\mathrm{Ir}(5\,\mathrm{nm})/\mathrm{Co}(t_{Co})$ trilayer samples with $t_{Co}$ = 2 and 6 nm were deposited directly on carbon-coated TEM grids with a 200-mesh grid size by sputtering under the same deposition parameters described earlier. The corresponding HRTEM images and SAED patterns for $t_{Co}$ = 2 and 6 nm are shown in Fig. \ref{fig:TEM}.

The HRTEM micrographs depicted in Fig. \ref{fig:TEM} (a) and (d) reveal densely packed nanocrystalline grains with well-defined grain boundaries distributed throughout the film. Lognormal fits to the grain-size distributions give median lateral sizes of $4.7 \pm 0.8$~nm for $t_{\mathrm{Co}} = 6$~nm and $3.3 \pm 0.2$~nm for 2~nm suggesting that the grain growth during sputter deposition is suppressed by decreasing the Co thickness. The crystallite sizes obtained from GIXRD are compared to the grain sizes obtained from HRTEM for the 2 and 6 nm Co-layer samples. The crystallite size obtained from GIXRD and the grain size obtained from HRTEM represent different structural length scales; GIXRD probes coherently diffracting domains, while HRTEM image analysis determines the dimensions of grains visible in the micrographs. Differences between the values obtained with these techniques are thus expected, and their comparison provides complementary insight into the nanoscale structure of the trilayers.
\begin{figure}[t!]
    \centering
    \includegraphics[width=1.0\linewidth]{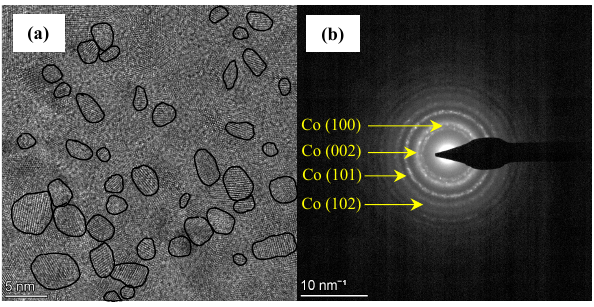}
    \includegraphics[width=1.0\linewidth]{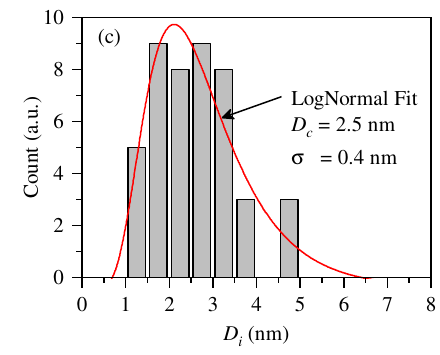}
    \caption{HRTEM images of the pure Co of a 5 nm thick film (a). The solid curves are drawn to show the crystalline boundaries. (b) is the corresponding SAED pattern of the HRTEM  images depicted in (a). (c) Distribution of the Co nanocrystallite sizes fitted with a lognormal distribution equation.}
    \label{fig:Co_TEM}
\end{figure}

The corresponding SAED patterns presented in Fig. \ref{fig:TEM}(b) and (e) consist of sharp concentric diffraction rings, characteristic of the polycrystalline nature of the samples, consistent with the GIXRD results. Upon indexing the diffraction rings, all the observed rings can be assigned to the FCC Ir phase, corresponding to the Ir(111), Ir(200), Ir(220), Ir(311), and Ir(331) crystallographic planes, in good agreement with JCPDS card No. 00-006-0598 for Ir \cite{Puthalath2026}. This does not imply that the Co layers are poorly crystalline. The strongest hcp Co reflections nearly coincide with Ir rings. Co(100) lies 2\% from Ir(111) and Co(101) 0.2\% from Ir(200), separations that cannot be resolved for rings broadened by nanometre-sized grains. Ir also scatters electrons considerably more strongly than Co, so the Ir rings dominate the pattern. The rings are weaker for the 2~nm sample, as expected from the smaller scattering volume. No separate oxide rings are resolved: the surface oxide identified by XPS is too thin to contribute appreciably, and the strongest CoO reflection, (200), lies within 2\% of Co(100).HRTEM and selected-area electron diffraction (SAED) measurements were used to investigate the microstructure and diffraction characteristics of a 5 nm pure Co film for comparison Fig. \ref{fig:Co_TEM}. SAED pattern provides complementary information on the crystalline nature of the Co film, whereas the HRTEM image provides information on the nanoscale grain structure.


The combined HRTEM and SAED images provide direct microscopic evidence for the high structural quality of the sputtered Co/Ir/Co trilayers. The agreement between the nanocrystalline grain morphology observed by HRTEM and the crystallite size obtained from GIXRD confirms that the structural evolution with decreasing Co thickness is governed primarily by the suppression of grain growth rather than by the formation of secondary phases. These structural characteristics are expected to modify the interface-to-volume ratio and, consequently, play an important role in determining the thickness-dependent magnetic behaviour discussed in the following sections.

\subsection{Magnetization Study}
The temperature-dependent magnetisation, $M(T)$, and field-dependent magnetisation, $M(H)$, measurements provide essential insights into the magnetic behaviour and thickness-dependent evolution of the magnetic properties in Co/Ir/Co trilayer heterostructures. Since Ir is non-magnetic, the observed magnetisation primarily originates from the Co layers, while the Co--Ir interfacial alloying can significantly modify their magnetic response. A 5-nm Ir spacer is far thicker than the sub-nanometre range over which Ir mediates interlayer exchange coupling between Co layers \cite{Parkin1991}. To the first order, two Co layers can therefore be treated as magnetically independent. They are not equivalent; however, the bottom layer grows on the HF-etched Si, whereas the top layer grows on polycrystalline Ir. In particular, $M(T)$, and $M(H)$ measurements enable identification of ferromagnetic (FM) states and determination of magnetic parameters, including the blocking temperature ($T_b$)~\cite{singh2025viable, Bedanta2009, Roy2023} , exchange bias field ($\Delta H_{EB}$), coercivity ($H_C$), saturation magnetization ($M_S$). Consequently, changes in $T_b$ with Co-layer thickness can be attributed to variations in Co-layer thickness and the local magnetic environment of Co nanocrystallites and Co-Ir interfacial alloying.

\begin{figure}[t]
    \centering
    \includegraphics[width=1\linewidth]{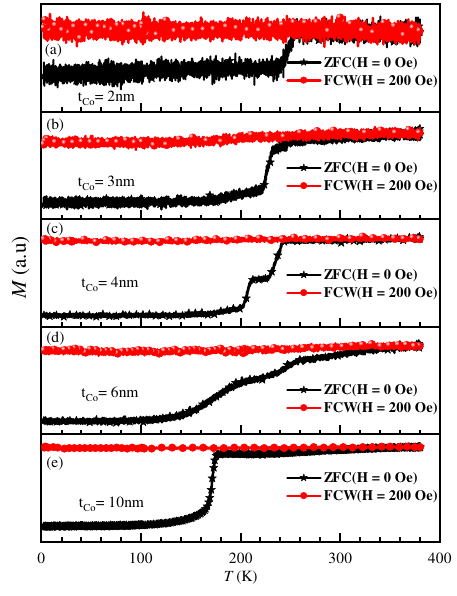}
    \captionsetup{font=small}
    \caption{ZFC-FC magnetisation data of the sample within the temperature range of 3 to 380K   }
    \label{fig:MT}
\end{figure}

\subsubsection{Temperature-dependent magnetisation study}\label{subsec3}
 To investigate the temperature-dependent magnetic response of the Co/Ir (5 nm)/Co trilayers, magnetisation measurements, $M(T)$, were performed in zero-field-cooled (ZFC) and field-cooled warming (FCW) modes over the temperature range of 3–380 K. The corresponding results are presented in Fig.~\ref{fig:MT}. For the ZFC protocol, the samples were cooled from $380$~K to $3$~K in the absence of an external magnetic field, whereas for the FCW protocol, the samples were cooled under an applied field of $200$~Oe, following a widely adopted measurement procedure reported in the literature \cite{singh2025viable, Knobel2008, Vigneswari2026Co}. The $M(T)$ data for both the branches were measured during warming under a constant in-plane measuring field ($H_m$) of 200~Oe. Inspection of the $M(T)$ curves in Fig.~\ref{fig:MT}(a)--(e) reveals three distinct magnetic characteristics: bifurcation of the ZFC and FCW magnetisation branches, which is temperature-dependent; a step increase in ZFC magnetisation around characteristic temperatures; and an irreversibility temperature, $T_{irr}$, below which the ZFC and FCW curves deviate, which is thickness-dependent.
 
All samples show qualitatively the same temperature-dependent magnetisation behaviour. The FCW magnetisation is almost temperature-independent over the measured temperature range. In contrast, the ZFC magnetisation is relatively low at low temperatures and increases in one or two distinct steps over narrow temperature ranges. These steps are observed at characteristic temperatures $(T_{\mathrm{b}})$, where the ZFC and FCW branches merge and are nearly coincident up to 380K. The ZFC curves do not show a distinct, pronounced maximum in the measured temperature range. The step is single and sharp for $t_{\mathrm{Co}}$ = 2, 3 and 10 nm and splits into two for $t_{\mathrm{Co}}$ =4 and 6 nm.


On warming, the ZFC magnetisation stays low and then rises in a narrow step onto the FCW branch; no maximum is observed below 380~K. The temperature-independent FCW branch and the open hysteresis loops that persist up to 380~K (Fig.~\ref{fig:MH}) rule out a significant superparamagnetic fraction. The step marks the temperature at which the switching field of the zero-field-cooled state falls to the order of the measuring field, causing the film to reverse its magnetisation direction.

\begin{figure}[ht]
    \centering
    \includegraphics[width=1\linewidth]{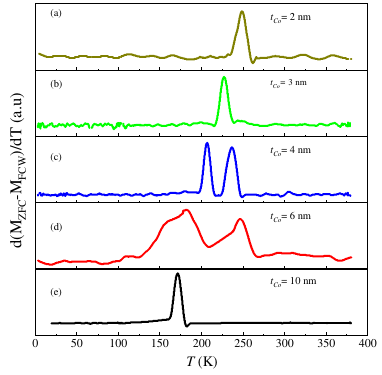}
    \caption{Derivative of difference between ZFC and FCW magnetisation vs temperature for the samples ${t_{Co}}$ = 2,3,4,6,10 nm (a)-(e)  }
    \label{fig:estTb}
\end{figure}

The blocking temperature, $T_{\mathrm{b}}$, was determined from the temperature at which the derivative of the difference between the zero-field-cooled and field-cooled-warming magnetisations, \cite{Mamiya2005}
\begin{equation}
    T_b = \operatorname*{argmax}_T \left[ \frac{d}{dT}{(M_{\text{ZFC}} - M_{\text{FCW}})} \right]
    \label{eq:Tb}
 \end{equation}
reaches its maximum. This derivative-based criterion was used to identify the temperature at which the irreversible component of the magnetisation changes most rapidly,\cite{Bruvera2015}. Following this approach, the corresponding results are presented in Fig.~\ref{fig:estTb}. We refer to it as the blocking temperature in the general sense of thermally activated reversal: below $T_{\mathrm{b}}$ the energy barriers to reversal in the measuring field cannot be overcome on the timescale of the measurement, and the configuration set during zero-field cooling is retained; above it the film reverses into the field direction. Unlike the blocking temperature of a superparamagnetic ensemble, $T_{\mathrm{b}}$ is not a property of isolated grains and does not scale with grain volume. It depends on the measuring field through $H_{\mathrm{C}}(T_{\mathrm{b}}) \approx H_{\mathrm{m}}$, and all values quoted here refer to $H_{\mathrm{m}} = 200$~Oe. One maximum is found for $t_{\mathrm{Co}} = 2$, 3 and 10~nm and two for 4 and 6 nm. $T_{\mathrm{b}}$ of the single-step samples and the lower $T_{\mathrm{b}}$ of the two-step samples decreases monotonically with Co thickness, from 248~K at 2~nm to 171~K at 10~nm.

\begin{figure}[b!]
    \centering
    \includegraphics[width=1\linewidth]{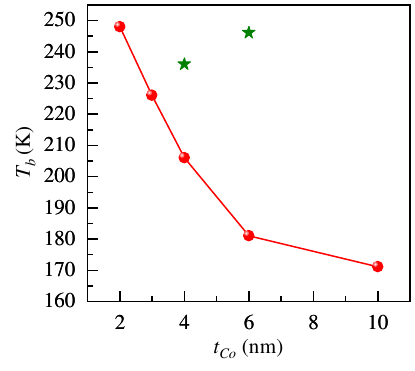}
    \captionsetup{font=small}
    \caption{Variation of the ${T_{b}}$ with ${t_{Co}}$, represented by the solid spheres. The green solid star symbols also represent ${T_{b}}$ for samples exhibiting two ZFC humps, with the higher-temperature hump corresponding to ${T_{b}}$ (samples with ${t_{Co}} =$ 4 and 6 nm in Fig.~\ref{fig:MT}(c) and (d), respectively).}
    \label{fig:Tbvstco}
\end{figure}

This behaviour reflects the thickness-dependent balance between Co/Ir interfacial and Co-volume contributions. This variation indicates a thickness-dependent balance between interfacial and volume contributions. From the data, samples with $t_{\mathrm{Co}}=4$ and $6$~nm exhibit two distinct peaks, whereas the other samples show a single peak. For the two-peak cases, the lower-temperature peak is considered as $T_{\mathrm{b}}$ and plotted as a function of $t_{\mathrm{Co}}$ in Fig.~\ref{fig:Tbvstco}, represented by solid spheres. The higher-temperature peaks are also included, denoted by solid star symbols.
The two steps for $t_{\mathrm{Co}}$ = 4 and 6~nm show that these films contain two populations with different switching fields. The natural candidates are the two Co layers. A 5~nm Ir spacer is far thicker than the sub-nanometre range over which Ir mediates appreciable interlayer exchange coupling \cite{Parkin1991,Ma2020}, so the layers are expected to switch independently, and their environments differ: the bottom layer sits on Si, where Co–Si intermixing produces interfacial disorder \cite{Roy2023}, while the top layer sits on Ir and carries a surface oxide. Interface disorder, grain
structure and surface oxidation all affect domain-wall pinning and the nucleation of reversal. Each layer therefore has its own $H_{\mathrm{C}}(T)$ and hence its own $T_{\mathrm{b}}$. For two steps to be resolved, the two layers must carry a significant moment and the switching temperatures must be separated by more than the width of each step. For $t_{\mathrm{Co}} = 10$~nm, the interfaces represent only a small fraction of each layer, the two layers behave similarly, and the steps merge. For $t_{\mathrm{Co}}=$ 2 nm a dead layer of about 0.7~nm per interface leaves little magnetic Co, especially in the bottom layer, so one of the steps may be too small to detect.


\subsubsection{Field-dependent magnetisation study}\label{subsec4}

\begin{figure*}[t!]
    \centering
    \includegraphics[width=1\linewidth]{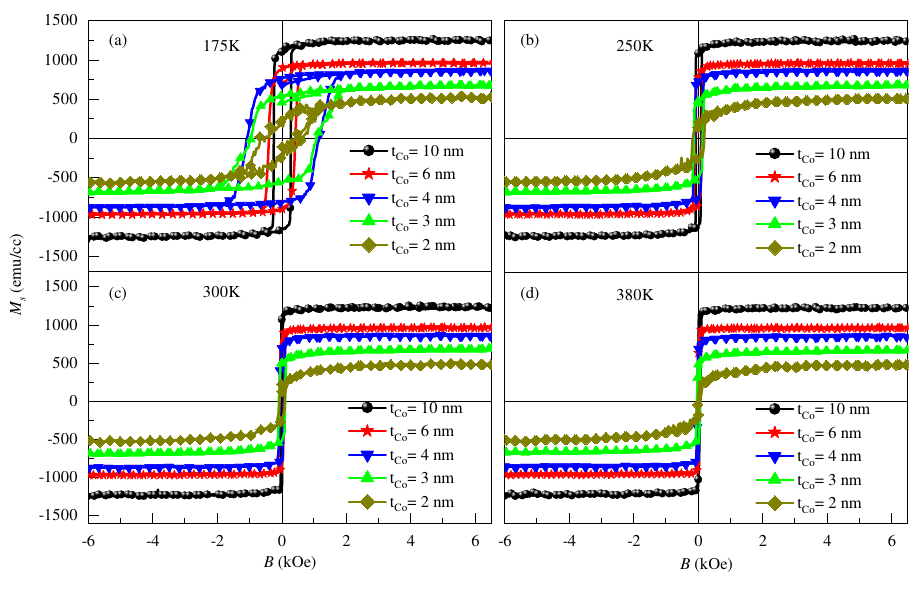}
    \caption{MH loops of the Co/Ir/Co trilayers for the $t_{Co}$= 2, 3, 4, 6, 10 nm }
    \label{fig:MH}
\end{figure*}

\begin{figure*}[htb]
    \centering
    \includegraphics[width=1\linewidth]{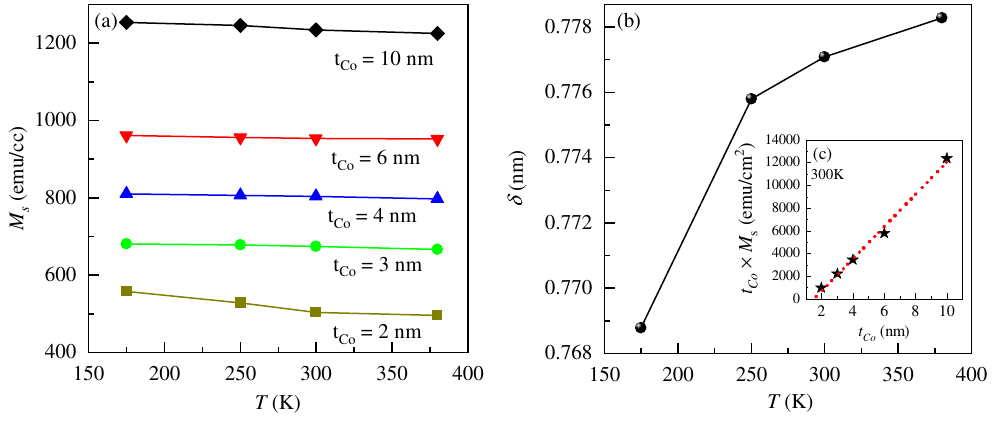}
    \caption{Variation of magnetic saturation with respect to the thickness of the cobalt layer }
    \label{fig:Ms}
\end{figure*}

In addition to acquiring a more profound understanding of the magnetic behaviour of the Co/Ir/Co trilayers, field-dependent magnetic studies were performed at 175 K, 250 K, 300 K, and 380 K. Fig. \ref{fig:MH} shows the corresponding MH loops for all Co thicknesses. The hysteresis loop exhibits very clear ferromagnetic behaviour over the temperature range, with the magnetic response correspondingly depending on both the temperature and the thickness of the Co layer. As the temperature increases, the loops are progressively narrower, suggesting a decrease in the magnetisation's resistance to field-driven reversal.

There is also a clear dependence on thickness. At certain temperatures, the intermediate-thickness samples show significantly greater coercive fields than the thinnest and thickest Co layers. The coercive field varies significantly with Co thickness. This suggests that the Co layer thickness and, consequently, the underlying magnetic energy distribution of the trilayers have a significant impact on the magnetisation reversal.

\begin{figure}[htb]
    \centering
    \includegraphics[width=1\linewidth]{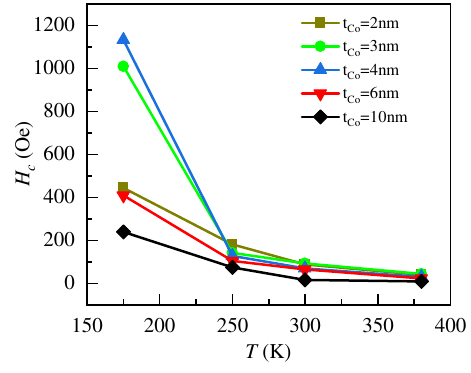}
    \caption{Variation of coercivity with respect to the temperature for $t_{Co}$ =2-10 nm }
    \label{fig:Hc}
\end{figure}

The temperature dependence of the coercivity $H_{C}$ extracted from the MH loops calculated using the Eq \ref{eq:coercivity},\cite{V2016} where ${H_{\mathrm{C}}^{R}}$ is the coercivity field on the right side of the loop and ${H_{\mathrm{C}}^{L}}$ is the coercivity field on the left side of the loop is shown in Fig.\ref{fig:Hc}. A pronounced decrease in coercivity with increasing temperature is observed in samples with different Co thicknesses, demonstrating the strong temperature dependence of the magnetic reversal field. In Hf/Co/Hf structures, Jafri et al. observed a non-monotonic dependence of coercivity on thickness with a maximum at an intermediate Co thickness of about 6 nm \cite{Jafri2022}. This behaviour qualitatively follows the increase in coercivity observed at intermediate Co-layer thicknesses in the present Co/Ir/Co trilayers. The reduction in $H_{C}$ indicates that thermal fluctuations significantly facilitate magnetisation reversal, thereby reducing the external field required to alter the magnetic state. Such temperature dependence is the sign of the thermally activated magnetisation reversal in magnetic systems with finite energy barriers.

\begin{equation}
H_{\mathrm{C}} =
\frac{\left|H_{\mathrm{C}}^{R}\right| +
\left|H_{\mathrm{C}}^{L}\right|}{2}
\label{eq:coercivity}
\end{equation}

The magnitude and temperature dependence of the coercivity strongly depend on $t_{Co}$. The maximum coercivity is observed in the 4 nm sample at lower temperature and the 2 nm sample shows comparatively lower coercivity. The 3 nm sample also has a higher coercive field, while the 6 and 10 nm samples have progressively lower values. The 3 nm sample also exhibits a higher coercive field, while the 6 and 10 nm samples show progressively lower values. With increasing temperature, the coercivity of all the samples decreases, and the difference between the thicknesses becomes progressively less pronounced. The resulting non-linear dependence of $H_{C}$ and $t_{Co}$ suggests that the magnetic reversal behaviour does not depend only on the magnetic volume of the Co layers\cite{Camarero2000}. Interestingly, 3 and 4 nm samples exhibit higher Co/Ir interfacial roughness as determined from XRR analysis. The increased interfacial roughness can cause local variations in magnetic anisotropy, thereby generating additional pinning sites that hinder magnetisation reversal and contribute to the enhanced coercivity observed at lower temperatures \cite{Zhao2001, Gautam2018}. This effect is expected to be more pronounced at low temperatures where thermal fluctuations are not sufficient to overcome readily the associated local energy barriers. With increasing temperature the influence of such pinning is reduced by thermally activated reversal, and consequently the coercivity decreases and the thickness dependence disappears. This is due to the combined effect of the effective magnetic anisotropy, Co/Ir interfacial effects, Co/Ir interface roughness \cite{Liu2000}, microstructural features and thermally activated reversal processes.

The temperature-dependent coercivity can be further related to the ZFC-FC magnetisation discussed above. The results demonstrate a significant thickness-dependent blocking temperature, with two distinct temperatures observed for the 4 and 6 nm samples. In parallel, \(H_{\mathrm{C}}\) continuously decreases with increasing temperature, showing that thermally assisted magnetisation reversal gets progressively easier at higher temperatures. The simultaneous thickness dependence of the blocking temperature and coercive field indicates that the thermally activated magnetic reversal behaviour evolves with $t_{Co}$. The coercive-field responses varied along the thickness series and showed a significant peak at low temperature in the 4 nm sample. This indicated that the magnetisation-reversal characteristics were dependent on the Co-layer thickness and were not only a function of the magnetic volume.

The saturation magnetisation, $M_{S}$, is obtained from the MH loops for the different temperature is shown in the Fig. \ref{fig:Ms}. For all the Co thicknesses, $M_{S}$ exhibits only very weak temperature dependence over the investigated range. However, a strong dependence on $t_{Co}$ is observed, $M_{S}$ increasing systematically with increasing Co layer thickness. At 175K $M_{S}$ increases approximately from 550 emu/cc for $t_{Co}$ = 2 nm to approximately 1250 emu/cc for $t_{Co}$ = 10 nm. The relatively weak variation of $M_{S}$ with the temperature is compared with the strong dependence of the Co thickness, indicating that the dominant variation of the measured magnetisation originates from the effective magnetic volume of the Co  \cite{Sharma2007}

Another parameter that can be evaluated from the magnetisation data is the magnetic dead layer thickness \cite{ElKhiraoui2008}. The magnetic dead layer is the magnetically inactive part of the Co layer and forms near the Co/Ir interface. To evaluate the contribution of the magnetic dead layer at the interface, the thickness dependence of $M_{S}$ was analysed using the conventional magnetic dead layer model eq \ref{eq:dead_layer}, \cite{Singh2022deadlayerstatic}
\begin{equation}
M_s=M_0\left(1-\frac{2\delta}{t_{\mathrm{Co}}}\right),
\label{eq:dead_layer}
\end{equation}
Where $M_{0}$ is the bulk cobalt value. From the fitting, it is calculated that $M_{0}$ value at 300K is 1422(emu/cc), which is consistent with the reported bulk value for hcp Co (1400--1450~emu/cc) \cite{Sharma2007, Jafri2022, Won2021}. In order to calculate the magnetic dead layer ($\delta$), $M_{S}$ $\times$ $t_{Co}$ is plotted against $t_{Co}$ and shown in the plot inside the Fig.\ref{fig:Ms} (c) for 300K. Similar fitting is performed for all the temperatures. The extracted value of the $\delta$ is plotted against temperature in Fig.\ref{fig:Ms} (b). A small but systematic increase in $\delta$ of $\sim$0.768 nm at 175K to 0.778 nm at 380K is observed. Although the variation is very small, it shows the weak temperature dependence of the magnetically inactive interfacial contribution. This behaviour suggests that the effective magnetic volume of the Co layer is slightly reduced when the temperature increases.

\begin{figure}[b]
    \centering
    \includegraphics[width=0.9\linewidth]{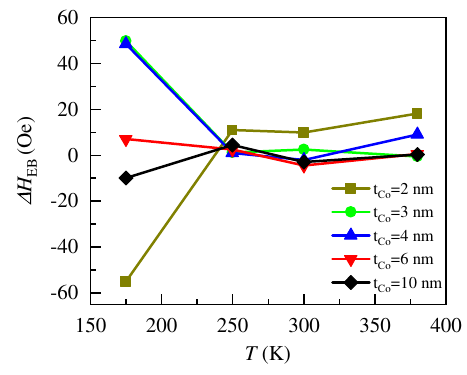}
    \caption{Exchange bias effect with respect to the temperature for $t_{Co}$ = 2-10 nm }
    \label{fig:H_EB}
\end{figure}

\begin{figure*}[t]
    \centering
    \includegraphics[width=1.0\linewidth]{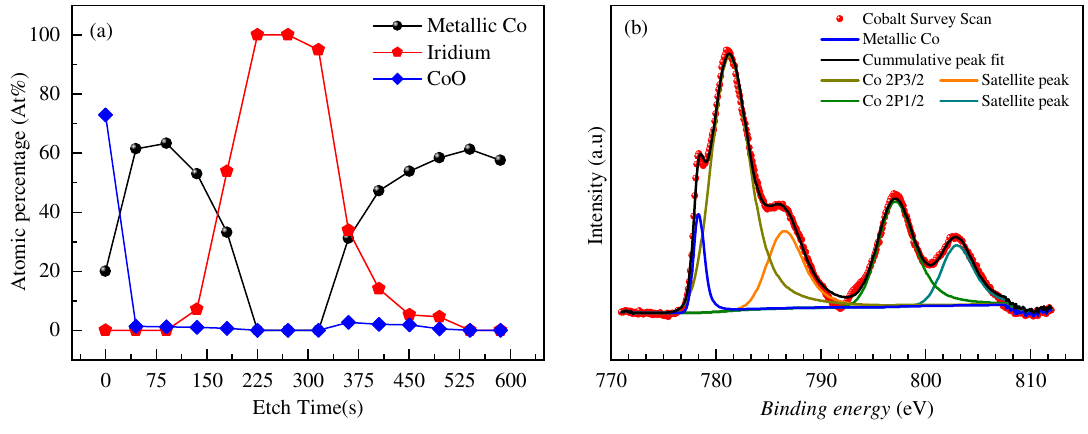}
    \caption{(a) Depth profile fitting of and (b) high-resolution Co 2p spectrum of $t_{Co}$ = 10 nm sample. }
    \label{fig:XPS}
\end{figure*}

Furthermore, exchange bias (EB) effect is observed in the $M(H)$  performed at different temperatures displayed in \ref{fig:MH}. The observed EB has been reported in bilayer and trilayer systems comprising FM/antiferromagnetic \cite{Ciprian2011, Ciprian2010}, FM/ferrimagnetic \cite{Mangin2003}, and FM/NM \cite{singh2020} heterostructures. In the present study, the hysteresis loops were measured after cooling the samples in the absence of an external applied magnetic field; therefore, the observed exchange-bias behaviour can be termed as the spontaneous exchange bias (SEB) reported elsewhere \cite{singh2020,Bufaial2024}. The observation of SEB in the present Co/Ir trilayer system is noteworthy and may be associated with intrinsic magnetic interactions within the multilayer structure or surface oxidation. In particular, we presume that oxidation of the top Co layer, resulting from exposure to ambient environmental conditions during sample handling and magnetic measurements, could contribute to the observed SEB behaviour. Such oxidation is plausible because the Co/Ir trilayers were not protected by a capping layer. The exchange-bias field ($\Delta H_{EB}$) is determined from the horizontal displacement of the hysteresis loops along the applied magnetic-field axis. It is calculated using the eq \ref{eq:exchange_bias}, \cite{V2016}. The temperature dependence of $\Delta H_{EB}$ is presented in Fig.~\ref{fig:H_EB}. Unlike the relatively systematic temperature dependence of the coercive field, $\Delta H_{EB}$ exhibits a pronounced dependence on both temperature and Co-layer thickness. Such spontaneous exchange-bias behaviour may be of potential interest for spintronic memory applications.

\begin{equation}
\Delta H_{\mathrm{EB}} =
\frac{\left|H_{\mathrm{C}}^{R}\right| -
\left|H_{\mathrm{C}}^{L}\right|}{2}
\label{eq:exchange_bias}
\end{equation}

It is very apparent from the Fig.~\ref{fig:H_EB} that $M(H)$ reveals a finite horizontal displacement of the magnetic loops in the thinner Co layers for $t_{Co}$ < 6 nm at temperatures below their characteristic blocking regime \cite{Meiklejohn1957,Ghoshani2023}. At 175 K, the $t_{\mathrm{Co}}$ = 3 and 4 nm samples exhibit relatively large positive $\Delta H_{\mathrm{EB}}$ values of approximately 50 Oe, whereas the $t_{\mathrm{Co}}$ = 2 nm sample shows a large negative value of about $-55$ Oe. In contrast, considerably smaller exchange-bias shifts are observed for the thicker Co layers.
With increasing temperature to 250 K, the magnitude of $ \Delta H_{\mathrm{EB}} $ decreases significantly for most samples and approaches to values close to zero, indicating a strong weakening of the exchange-bias effect. At 300 K $\Delta H_{\mathrm{EB}}$ is very small, typically of order $\pm10$ Oe. 

Further, to investigate the possible origin of the EB associated with the presence of the surface oxidation, XPS depth profiling was performed on the sample with $t_{\mathrm{Co}}$ = 10 nm, as shown in Fig.\ref{fig:XPS}(a). As seen from the data, at time stamp $t=0 s$ a strong CoO contribution is detected at the initial surface that rapidly decreases with sputtering depth as time proceeds, whereas the metallic Co contribution slowly increases. The metallic Co signal then decreases upon reaching the Ir layer and increases again at greater depth, corresponding to the bottom Co layer. Depth-dependent evolution confirms that oxidation occurs mostly on the exposed top Co surface, while the underlying Co is largely metallic.  The formation of this oxidised Co region is consistent with a contribution to the observed EB through exchange coupling between the oxidised and metallic Co regions. The oxide is therefore confined to the top surface and is at most a few \AA{} thick. Because Ar$^{+}$ sputtering reduces Co oxides preferentially, this profile gives a lower bound on the oxide content. The survey scan of the top Co layer is also presented in Fig.\ref{fig:XPS} (b). The presence of the Co 2p spectrum shown in Fig.\ref{fig:XPS} (b) provides complementary evidence for the coexistence of metallic Co and oxidised Co environments. The measured spectrum exhibits five distinct peaks at binding energies of 778.1 eV, 780 eV, 786 eV, 796 eV, and 802 eV, assigned to metallic Co, oxide-related components, and characteristic satellite features. The combined depth-dependent composition of Co 2p spectral analysis, therefore, establishes the presence of a Co-oxide-rich region at the exposed Co surface overlying metallic Co,\cite{Vigneswari2026Co, Biesinger2011, Stienen2021}. The oxidised Co-rich surface layer identified by XPS is consistent with a CoO-like phase, which can act as an antiferromagnetic layer and form an exchange-coupled interface with the underlying ferromagnetic metallic Co,\cite{Meiklejohn1956, Meiklejohn1957, Demirci2011}. Uncompensated antiferromagnetic moments at this interface can interact with the ferromagnetic Co moments, introducing a unidirectional contribution to the magnetic anisotropy and resulting in a horizontal displacement of the hysteresis loop \cite{Tomou2006}. Thus, the XPS results provide the direct chemical evidence for the oxidised Co component that can act as the antiferromagnetic pinning region associated with the exchange bias behaviour,\cite{Gnoli2024}. 

The magnitude of this interfacial pinning is expected to depend strongly on the structural and magnetic state of the ultrathin Co layers. In particular, the comparatively larger interfacial roughness observed for the 3 and 4 nm (see ~ Fig.~\ref{fig:I}) samples may produce spatial variation in the local magnetic energy landscape and the density of pinning sites that impede magnetisation reversal. This provides a possible explanation for their enhanced coercivity and measurable SEB fields at low temperature below $T_{b}$,\cite{Kumar2016}. Upon increasing the temperature beyond the blocking temperature, exchange bias becomes very small,\cite{Bundela2022}. The negative $\Delta H_{EB}$ observed for the 2 nm sample indicates that the net interfacial exchange configuration differs in the ultrathin limit. The sign reversal should not be solely due to the geometrical position of an antiferromagnetic layer relative to the ferromagnetic layer. Instead, it can arise from a change in the relative contribution of exchange-coupled uncompensated or pinned spins, their preferred orientation after field cooling and the reversal behaviour of the ferromagnetic layer \cite{Meiklejohn1957}. In a 2 nm sample, the reduced magnetically active Co thickness, together with oxidised and magnetically inactive interfacial regions, can modify the relative weight of competing interfacial exchange contributions \cite{Demirci2011}. The Negative value of $\Delta H_{EB}$ therefore suggests a distinct net interfacial spin configuration in the ultra-thin limit, rather than a different oxide phase being established solely from the sign of the loop shift\cite{Heinrich2008}.

\section{Conclusion}\label{sec13}

In summary, this work presents a systematic investigation of the influence of Co-layer thickness on the structural, microstructural, and temperature-dependent magnetic properties of sputter-deposited Co/Ir/Co trilayers, with a constant 5 nm Ir spacer. The trilayers have a nanocrystalline structure with thickness-dependent crystallite size and interfacial morphology. The saturation magnetisation increases with Co thickness, and the thickness-dependent study results in an effective magnetic dead-layer thickness of about 0.77 nm. The ZFC magnetisation jumps in a narrow step onto the FCW branch at a blocking temperature that decreases from 248 to 171 K with increasing $t_{\mathrm{Co}}$. The step occurs where the switching field falls to the order of the 200 Oe measuring field. It corresponds to the thermally activated reversal of continuous ferromagnetic layers. The 4 and 6 nm samples exhibit two-step magnetisation reversal, suggesting distinct switching processes within the Co layers. The coercivity exhibits the non-monotonic dependence on Co layer thickness with enhanced values for 3 and 4 nm Co layers in accordance with the effect of interfacial roughness and magnetic pinning. The coercivity decreases gradually with temperature, suggesting increasing thermal activation during magnetisation reversal.  Low-temperature hysteresis measurements show a thickness-dependent exchange-bias response, with opposing indications of $\Delta H_{EB}$ for 2 and 3-4 nm Co films. Overall, these findings provide a systematic understanding of the relationship between Co thickness, interfacial structure, and temperature-dependent magnetic behaviour in Co/Ir/Co trilayers.

\bmhead{Acknowledgements}

The authors SSS, DS, and KRK sincerely acknowledge the financial assistance provided by the Science and Engineering Research Board (SERB), Department of Science and Technology (DST), Government of India, through Grant No.~EEQ/2022/000354. The authors also acknowledge the Central Instrumentation Facility (CIF), IIT Gandhinagar, for providing access to the facilities for various material characterisations. Financial support from DST-FIST under Grant No.~SR/FST/PS-I/2020/147 (C) is gratefully acknowledged for carrying out the magnetisation measurements using the Physical Property Measurement System (PPMS) at the Cryogen-Free Measurement Facility, IIT Gandhinagar.

\section*{Declarations}

The authors declare that they have no known competing financial interests or personal relationships that could have appeared to influence the work reported in this paper.

\bibliography{sn-bibliography}

\end{document}